\documentclass[showpacs,aps,prd,twocolumn,nofootinbib,11]{revtex4}
\usepackage{amsmath}
\usepackage{epsfig}
\usepackage{subfigure}
\usepackage{float}
\usepackage{longtable}
\usepackage{times}
\usepackage{txfonts}
\usepackage{color}
\usepackage{ulem}
\newcommand{\ep}{\varepsilon}

\newcommand{\eqs}[1]{\begin{equation} \begin{split} #1\end{split} \end{equation} }

\newcommand{\ie}{{\it i.e.}}

\newcommand{\etal}{{\it et al.}}
\newcommand{\Br}{{\rm Br}}
\newcommand{\Q}{{\cal Q}}

\newcommand{\cf}[1]{{Fig.~\ref{#1}}}

\newcommand{\beq}[1]{
\begin{equation}\label{#1}}
\newcommand{\eeq}{\end{equation}}
\newcommand{\bea}[1]{
\begin{eqnarray}\label{#1}}
\newcommand{\eea}{\end{eqnarray}}
\newcommand{\out}{\raise-3pt\hbox{\scriptsize    out}}

\usepackage{ifpdf}
\ifpdf
\usepackage[pdftex]{hyperref}
\else
\usepackage[hypertex]{hyperref}
\fi

\hypersetup{
  pdftitle={},%
  pdfauthor={},%
  pdfsubject={},%
  pdfkeywords={},%
  pdfstartview={},%
  bookmarksopen=true, breaklinks=true, debug=true, %
  colorlinks=true, linkcolor=red, citecolor=blue, urlcolor=blue
}

\begin{document}

\title{Heavy-Quarkonium Production in High Energy Proton-Proton Collisions at RHIC}

\author
{Stanley J. Brodsky$^a$ and Jean-Philippe Lansberg$^{a,b}$\protect\footnote{Present address at Ecole Polytechnique.}
}

\preprint{SLAC-PUB-13736,INT-PUB-09-037}

\affiliation{$^a$ SLAC National Accelerator Laboratory, Theoretical Physics, Stanford University, Menlo Park, CA 94025, USA\\
$^b$Centre de Physique Th\'eorique, \'Ecole Polytechnique, CNRS,  91128 Palaiseau, France
}

\begin{abstract}
We update the study of the total $\psi$ and $\Upsilon$ production cross section in proton-proton collisions at  
RHIC energies using the QCD-based Color-Singlet (CS) Model, including next-to-leading order partonic matrix elements.
We also include charm-quark initiated processes which appear at  leading order  in $\alpha_s$, but which have so 
far been overlooked in such studies.   Contrary to earlier claims,  we show that the CS yield is consistent 
with measurements over a broad range of $J/\psi$ rapidities.   We also find that charm-quark initiated processes, 
including both intrinsic and sea-like charm components,  typically contribute at least  20 $\%$ of the direct $J/\psi$ 
yield, improving the agreement with  data both for the integrated cross section and its rapidity dependence.  The key 
signature for such  processes is the observation of a charm-quark jet opposite in azimuthal angle $\phi$ 
to the detected $J/\psi.$   Our results have impact on the proper interpretation of heavy-quarkonium production in 
heavy-ion collisions and its use as a probe for the  quark-gluon  plasma.
\end{abstract}
\maketitle

The hadroproduction of 
$J/\psi$ and $\Upsilon$   is one of the key topics in 
phenomenological QCD.  As opposed to lighter mesons, it is a priori straightforward to compute their production
rates from gluon-induced subprocesses such as 
$ g g \to \Q g$  (\cf{diagrams} (a)), particularly since one can use nonrelativistic approximations.
However, there are many outstanding theoretical issues, including the role of 
color-octet (CO) states,
the impact of next-to-leading order (NLO) -- and even 
higher order -- QCD corrections (\cf{diagrams} (c,d)), and the role of hard subprocesses such as $g c \to J/\psi c$ 
(\cf{diagrams} (b)) which utilize the $c$-quark distribution in the proton. 
Other  issues include the $J/\psi$
polarization puzzle, the factorization-breaking strong nuclear dependence 
in $J/\psi$ hadroproduction at high $x_F$, and the uncertain effects of rescattering and energy loss mechanisms.  
All of these issues have impact on the proper interpretation of heavy-quarkonium production in heavy-ion collisions 
and its use as a probe for the  quark-gluon  plasma. For recent reviews, see~\cite{reviews}.

It is widely accepted that $\alpha^4_s$ and $\alpha^5_s$ corrections to the CSM~\cite{CSM_hadron} are fundamental for
understanding the $p_T$ spectrum of $J/\psi$ and $\Upsilon$ produced in
high-energy hadron  collisions~\cite{Campbell:2007ws,Gong:2008sn,Artoisenet:2008fc,procNNLO,Li:2008ym,Lansberg:2009db}. 
However, if anomalously large contributions
to the total cross section arise from $\alpha^4_S$ graphs, this would cast doubt on the convergence
of the expansion in $\alpha_s$.  It is thus important to check that LO and NLO
predictions are close to each other and in agreement with
experimental data.
In this paper we carry out the first theoretical analysis in the CSM at NLO accuracy of
the total $J/\psi,$  $\psi(2S),$  and  $\Upsilon$ production in $pp$ collisions at the BNL RHIC.
We show that hard subprocesses based on CS $Q \bar Q$ configurations alone are
sufficient to account for the observed magnitude of the $p_T$-integrated
cross section. In particular, the predictions at
LO~\cite{CSM_hadron} and NLO~\cite{Campbell:2007ws,Gong:2008sn} accuracy
are both compatible  with measurements by the PHENIX collaboration at RHIC~\cite{Adare:2006kf} within present errors.
We shall also show that hard subprocesses involving the charm
quark distribution of the colliding protons (\cf{diagrams} (b)) which constitute part of the LO ($\alpha_s^3$) rate,
are responsible for a significant fraction of the observed yield.
Reactions such as $g c \to J/\psi c$  (thereafter referred to as $cg$ fusion) 
also produce  a charm jet opposite in azimuthal angle
to the  $J/\psi$; 
furthermore, the rapidity dependence of this ``away-side"
correlation is strongly sensitive to  the mechanism for the creation of the $c$-quark in the proton. 
An analysis
of the invariant mass distribution of the $J/\psi+D$ pair may also shed 
light on possible  contributions beyond the color singlet model, as described by the Color Transfer
Mechanism (CTM)~\cite{Nayak:2007mb,Nayak:2007zb}.

\begin{figure}[t!]
\centering
\subfigure[]{\includegraphics[scale=.39]{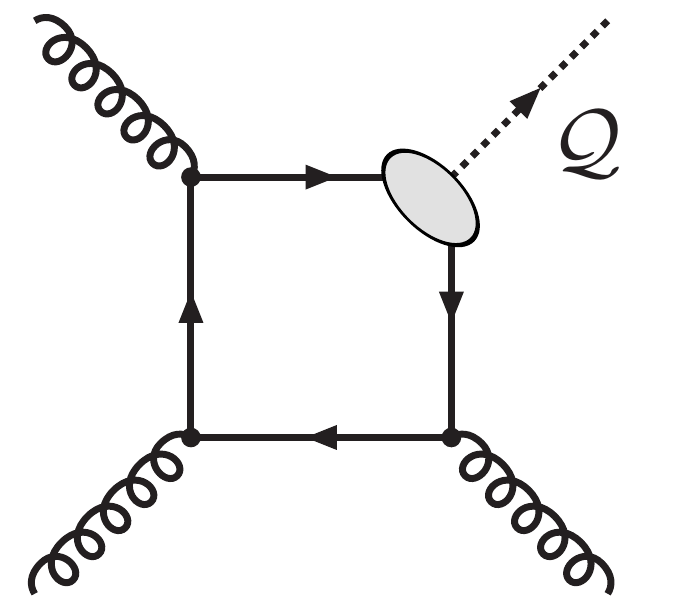}}\hspace*{-.2cm}
\subfigure[]{\includegraphics[scale=.39]{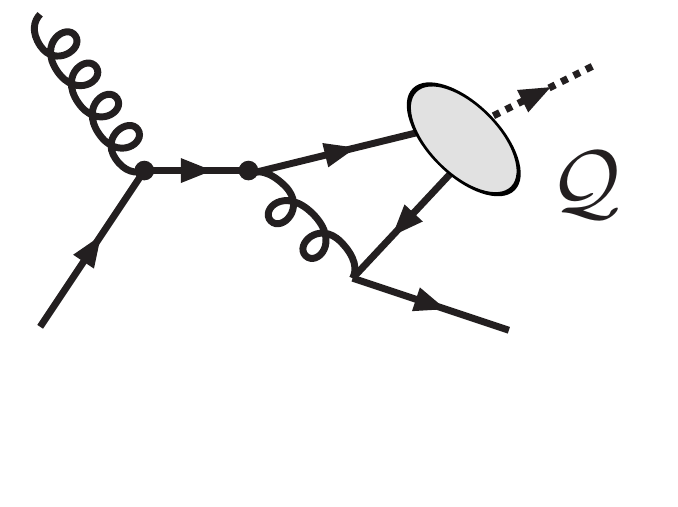}}\hspace*{-.2cm}
\subfigure[]{\includegraphics[scale=.39]{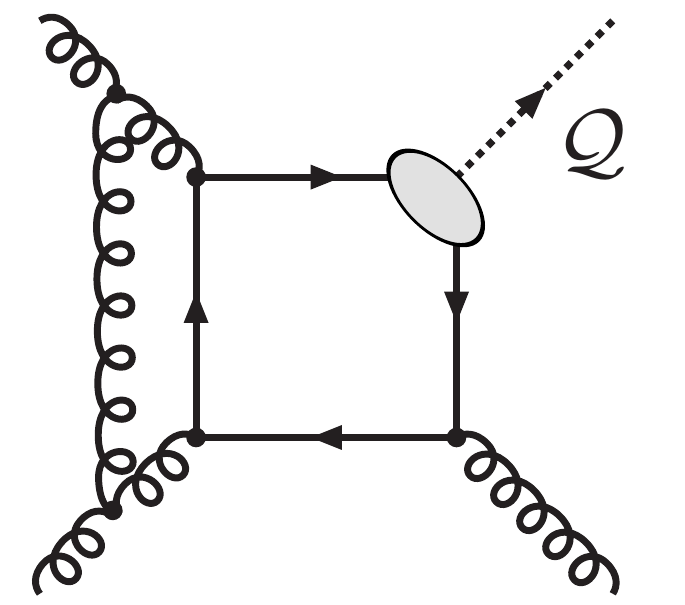}}\hspace*{-.2cm}\\
\subfigure[]{\includegraphics[scale=.39]{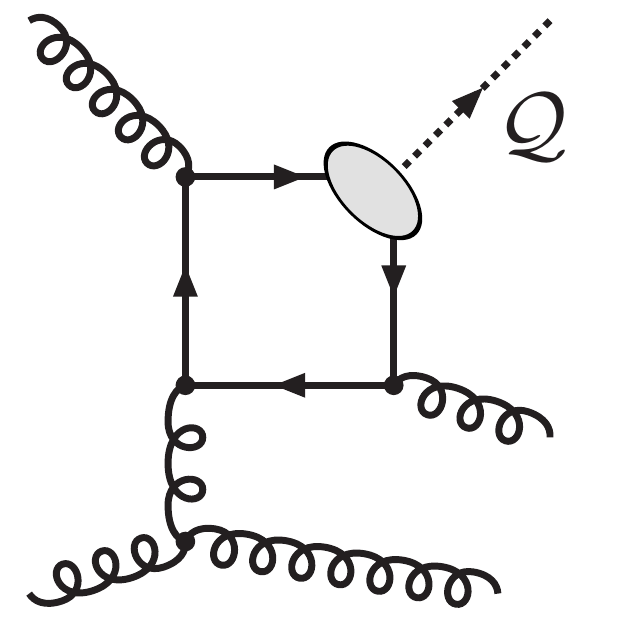}}\hspace*{-.2cm}
\subfigure[]{\includegraphics[scale=.39]{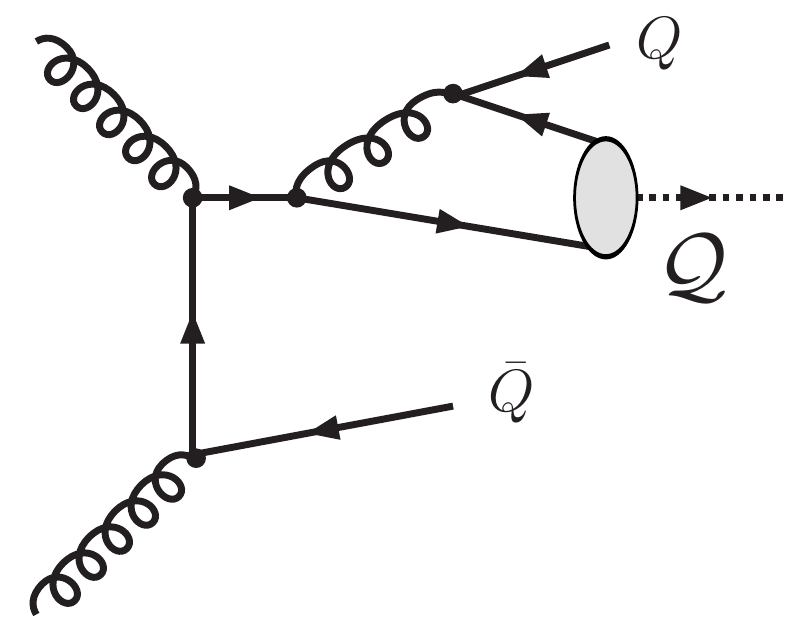}}\hspace*{-.2cm}
\subfigure[]{\includegraphics[scale=.39]{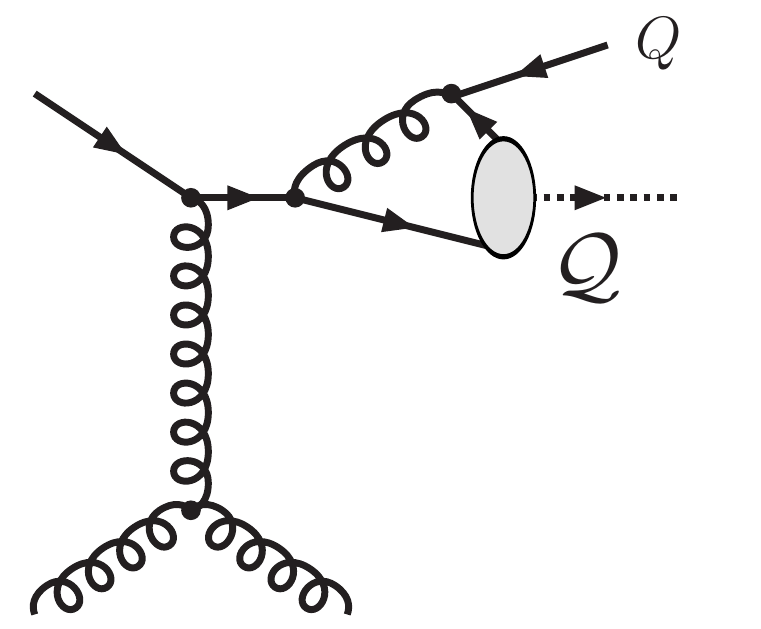}}
\caption{Representative diagrams contributing to $^3S_1$ quarkonium {(denoted $\Q$)} hadroproduction via
CS channels at orders $\alpha_S^3$ (a,b), $\alpha_S^4$ (c,d,e,f).
The quark and antiquark attached to the ellipsis are taken as on-shell
and their relative velocity v is set to zero.}
\label{diagrams}
\end{figure}

Subprocesses involving $cg$ fusion with a  charm quark from the proton have been considered
in ~\cite{Martynenko:1994kk,Qiao:2003pu}
with the main focus on the high $p_T$ spectrum
At low $p_T$,
the typical scale of the production process is rather small, and thus one does not expect higher-order QCD corrections such as gluon
splitting into $c \bar c$ to give a significant contribution to the total cross section
For  example, the contribution
to the total cross section from the
process $gg\to J/\psi c \bar c$, appearing at $\alpha_s^4$ (\cf{diagrams} (e))~\cite{Artoisenet:2007xi},  
is at the level of 0.5 $\%$.   In contrast, in the case of  intrinsic charm (IC) contributions, the $c$  and $\bar c$ quarks are created from  two soft gluons  connecting to different valence quarks in the proton as in the BHPS model~\cite{Brodsky:1980pb};  such contributions are relevant to charmonium production at all scales. The contribution from 
$c\gamma$ fusion was studied in  photoproduction in~\cite{Berger:1982fh}.

We shall focus here on the ``direct" hadroproduction of the $J/\psi$, $\psi(2S),$ and $\Upsilon(1S)$ without the contribution arising from the decay of heavier states; this avoids
the discussion of the production mechanisms of $P$-waves which are not well understood.  Although the total cross section for $L=1$  states has been studied at NLO~\cite{Petrelli:1997ge}, an
effective evaluation of the production cross section requires the introduction of an infrared cut-off (as for their
decay~\cite{barbieri}) or CO contributions~\cite{Bodwin:1994jh}  which introduce new unknown
non-perturbative parameters.   Furthermore, the impact of the off-shellness of initial gluon
on the $\chi_{c1}$ yield may be significant~\cite{Hagler:2000dd,Baranov:2007dw}. We have also restricted 
our analysis to the integrated-$p_T$ distribution. Indeed, as noticed at the Tevatron energy~\cite{Campbell:2007ws,Gong:2008sn},
the NLO $p_T$ distribution, contrary to the integrated one, can be negative at low $p_T$. 
In addition, initial-state radiation~\cite{Berger:2004cc} would also be expected to  significantly modify  the spectrum at small $p_T$ and to increase $\langle p_T^2 \rangle$.

In the case of $J/\psi$ hadroproduction, the PHENIX data~\cite{Adare:2006kf} includes the direct yield, 
but also
a $B$ feed-down fraction ($4 ^{+3}_{-2}\%$~\cite{Oda:2008zz}), a $\psi(2S)$ feed-down 
($8.6 \pm 2.5\%$ for $|y| <0.35$) and a $\chi_c$ feed-down  estimated to be $<42\%$  at $90\%$ C.L.~\cite{Oda:2008zz}.
A recent analysis~\cite{Faccioli:2008ir} from fixed-target
measurements in $pA$ suggests that it amounts to $25 \pm 5 \%$, while the CDF measurement in $pp$ at Fermilab
gives $30 \pm 6\%$ of the {\it prompt} yield for $p_T> 4$ GeV~\cite{Abe:1997yz}. For our analysis, we will make the
hypothesis that the $\chi_c$ feed-down fraction is $30 \pm 10\%$ of the prompt yield independent of rapidity.
Overall, we shall take {$F^{\rm direct}_{J/\psi}= 59 \pm 10 \%$ and multiply the PHENIX results by this factor}. The differential $J/\psi$ production cross section vs 
$y$ has been measured 
by PHENIX in the central ($|y|<0.35$) as well as in the forward ($1.2<|y|<2.2$) regions~\cite{Adare:2006kf,daSilva:2009yy}.
The extrapolation to the direct yield using $F^{\rm direct}_{J/\psi}= 59 \pm 10 \%$
is shown on \cf{fig:xsection} (a).
For the $\psi(2S)$, only a negligible $B$ feed-down competes with the direct mechanism. The preliminary measurement by PHENIX
is shown on \cf{fig:xsection} (b).
The $\Upsilon(1S+2S+3S)$ cross section  has been
measured by STAR~\cite{Djawotho:2007mj} and PHENIX~\cite{daSilva:2009yy} in the central region,
and by PHENIX~\cite{Xie:2005ac} in the forward regions. 
From the CDF analysis~\cite{Affolder:1999wm} at $p_T > 8$ GeV,
$ 50\%$ of the $\Upsilon(1S)$ are expected to be direct. Using the relative yields from \cite{Acosta:2001gv}, we expect
$42 \pm 10\%$ of the $\Upsilon(1S+2S+3S)$ signal to be direct $\Upsilon(1S)$. PHENIX and STAR data {mutiplied by}
 this  fraction are displayed on \cf{fig:xsection} (c).

In the CSM~\cite{CSM_hadron}, the matrix element to create
a $^3S_1$
 quarkonium ${\Q}$ of momentum $P$ and polarisation $\lambda$
 accompanied by
other partons, noted $j$, is the product of the amplitude to create
the corresponding heavy-quark pair, a spin
 projector $N(\lambda| s_1,s_2)$ and
$R(0)$, the radial wave function at the origin in the configuration
space, obtained from the leptonic width~\cite{Amsler:2008zzb}, namely 
\eqs{ \label{CSMderiv3}
{\cal M}&(ab \to {\Q}^\lambda(P)+j)=\!\sum_{s_1,s_2,i,i'}\!\!\frac{N(\lambda| s_1,s_2)}{ \sqrt{m_Q}} \frac{\delta^{ii'}}{\sqrt{N_c}} 
\frac{R(0)}{\sqrt{4 \pi}}\\\times&
{\cal M}(ab \to Q^{s_1}_i \bar Q^{s_2}_{i'}(\mathbf{p}=\mathbf{0}) + j)
}
where $P=p_Q+p_{\bar Q}$, $p=(p_Q-p_{\bar Q})/2$, 
$s_1$,$s_2$ are the heavy-quark spin and $\delta^{ii'}/\sqrt{N_c}$ is the projector onto a CS state.
In the non-relativistic limit, $N(\lambda| s_1,s_2)$
can be written as 
$ \frac{\ep^\lambda_{\mu} }{2 \sqrt{2} m_Q } \bar{v} (\frac{\mathbf{P}}{2},s_2) \gamma^\mu u (\frac{\mathbf{P}}{2},s_1) \,\, $
where $\ep^\lambda_{\mu}$ is the polarisation vector of the quarkonium. The sum over the spins yields to traces
evaluated in a standard way.

In our evaluation, we use the partonic matrix elements from Campbell,
Maltoni and Tramontano~\cite{Campbell:2007ws} to compute the
LO and NLO cross sections from gluon-gluon and light-quark gluon fusion. We guide the reader to~\cite{Campbell:2007ws}
for details concerning the derivation of ${\cal M}(ab \to {\Q}^\lambda(P)+j)$ at $\alpha_s^4$, the corresponding 
expressions at $\alpha_s^3$ can be found in~\cite{Baier:1983va}.
In the case of the $cg$ fusion {(at LO)}, we use the framework described in~\cite{Artoisenet:2007qm} based on the
tree-level matrix element generator {\small MADONIA}~\cite{Madonia}.
For the parameters entering the cross section evaluation, we have taken $|R_{J/\psi}(0)|^2=1.01$ GeV$^3$ and
$|R_{\psi(2S)}(0)|^2=0.639$ GeV$^3$.
We also take Br$(J/\psi \to \ell^+\ell^-)=0.0594$
and  Br$(\psi(2S) \to \ell^+\ell^-)=0.0075$.  For the $\Upsilon(1S)$, we will choose $|R(0)|^2=7.6$ GeV$^3$,
and Br$(\Upsilon\to \ell^+\ell^-)=0.0218$. The uncertainty bands for the resulting predictions are obtained from the {\it combined}
 variations of the heavy-quark
mass within the ranges $m_c=1.5\pm 0.1$ GeV and
$m_b=4.75 \pm 0.25$ GeV,\footnote{{It is common to see a wider range used for $m_c$ in NLO evaluations of 
open-charm cross sections, \ie~$m_c=1.5\pm 0.2$ GeV (see~\cite{Cacciari:2005rk}). In the case quarkonium production
within the CSM, such values so different from $M_{\Q}/2$ may require the inclusion of non-static corrections, which is beyond the scope of our analysis. See also our comment regarding the $\psi(2S)$ results.}} the
factorization $\mu_F$ and the renormalization $\mu_R$ scales
 chosen\footnote{In principle, the renormalization scale ambiguity can be removed using the method described
  in \cite{Binger:2006sj}.}
in the couples $((0.75,0.75);(1,1);(1,2);(2,1);(2,2))\times m_T$ with $m^2_T=4m_Q^2+p_T^2$.
Neglecting relativistic corrections, one has in the CSM, $M_{J/\psi}=M_{\psi(2S)}=2m_c$ and
$M_{\Upsilon}=2m_b$. {The parton distribution used was the LO set {\small CTEQ6\_L}~\cite{Pumplin:2002vw} for the LO $gg$ fusion, 
the NLO set {\small CTEQ6\_M} for
the $gg+gq$ NLO one and, for the $cg$ fusion, the LO set {\small CTEQ6.5c}~\cite{Pumplin:2007wg} based on a recent global PDF fit
including IC}. We have employed three choices for the charm distribution: (i) without IC [$c(x,\mu_0)=0$ ($\mu_0=$1.2 GeV)], (ii) with  BHPS IC~\cite{Brodsky:1980pb}
($\langle x \rangle_{c+\bar c}{\equiv\int^1_0  x [c(x)+\bar c(x)] dx=}2\%$) and (iii) with sea-like IC ($\langle x \rangle_{c+\bar c}=2.4\%$) . While there does exist an intrinsic $b$-quark content in the proton scaled by $m^2_c/m^2_b$  relative to IC,  its corresponding contribution to $\Upsilon+b$ is additionally suppressed at RHIC energy by phase space due to the presence of an additional $b$-quark in the final state.

\begin{figure}[!ht]
{\includegraphics[width=.99\columnwidth]{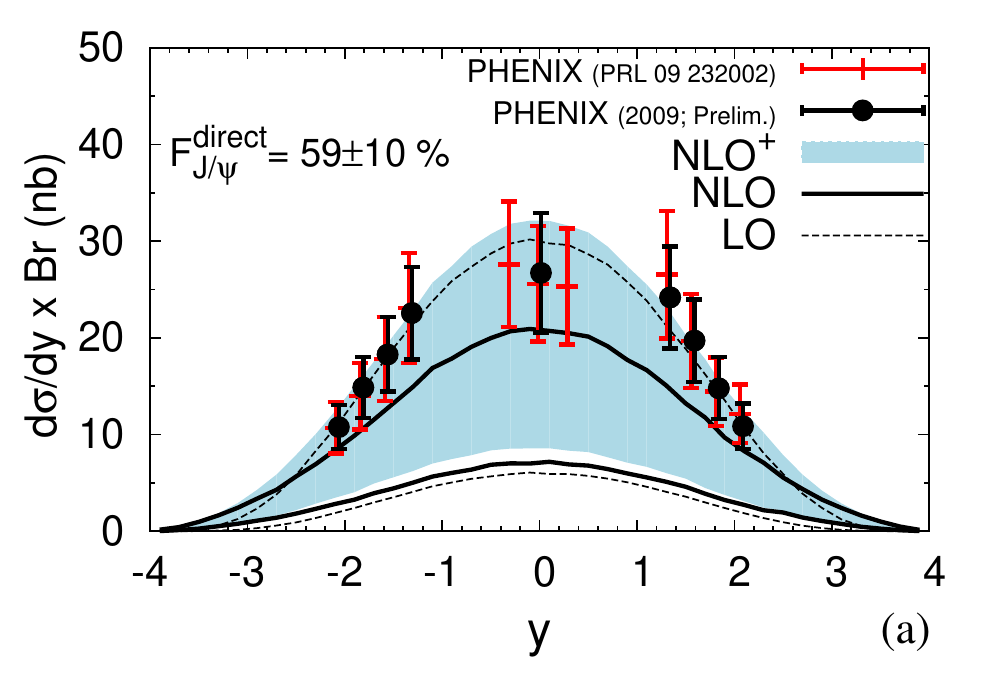}}\\%\hspace*{-.4cm}
{\includegraphics[width=.99\columnwidth]{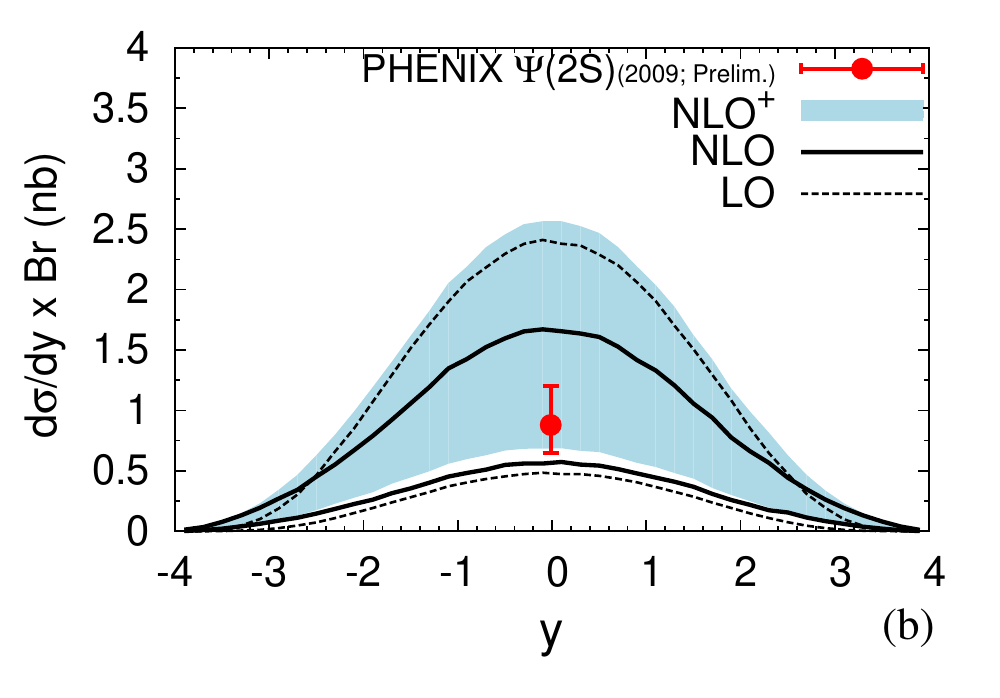}}\\%\hspace*{-.4cm}
{\includegraphics[width=.99\columnwidth]{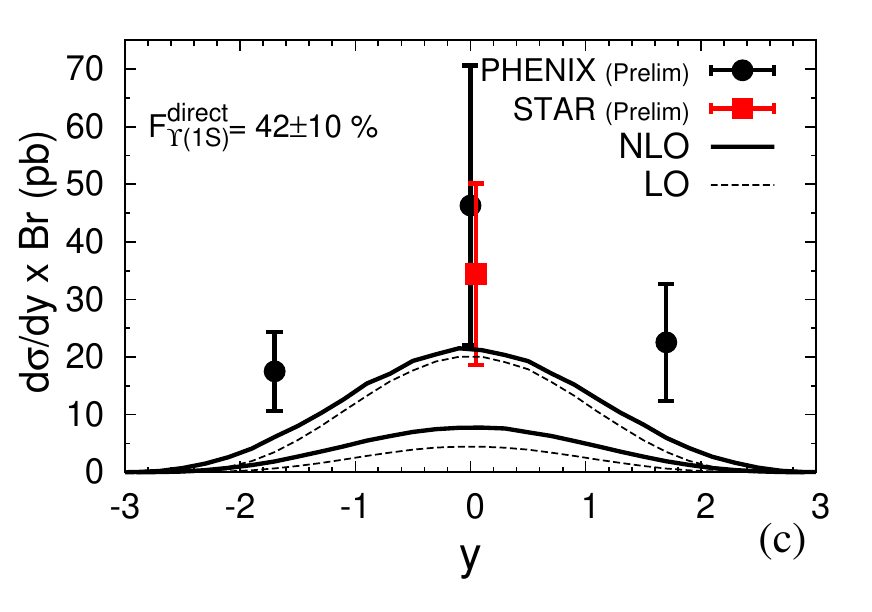}}\\\vspace*{-.5cm}%}
\caption{(a) {$d\sigma^{direct}_{J/\psi}/dy\times \Br$ from the
measurements by PHENIX~\cite{Adare:2006kf,daSilva:2009yy} multiplied by our estimate of $F^{\rm direct}_{J/\psi}$}
compared to the CSM at LO ($\alpha^3_s$) by $gg$ fusion 
only (thin-dashed lines),  at NLO (up to $\alpha_s^4$) by $gg$ and $qg$ fusion only (thick-solid lines)
and the sum ``NLO + $cg$ fusion'' with the sea-like $c(x)$~\protect\cite{Pumplin:2007wg}, denoted NLO$^+$ (light-blue band).
(b) same as (a) for the $\psi(2S)$ with PHENIX data~\cite{daSilva:2009yy}. (c)
same as (a) for the direct $\Upsilon$ with STAR~\cite{Djawotho:2007mj} and PHENIX~\cite{Xie:2005ac,daSilva:2009yy} 
preliminary measurements for $\Upsilon(1S+2S+3S)$ {multiplied by our estimate of $F^{\rm direct}_{\Upsilon}$} 
(without NLO$^+$, see text).
{The gaps between the two solid and the two dashed lines as well as the band reflect the variation of the cross section after
a {\it combined} variation of the scales and the masses as indicated in the text.}
}
\label{fig:xsection}
\end{figure}

We now describe our results.  As shown in \cf{fig:xsection} (a) and (b), the
yields at LO and NLO accuracy are consistent in size, and the uncertainty of the latter one 
(indicated by the two curves in both cases) is smaller than that of the LO. This provides  some indication that we are in a proper perturbative regime.
The yields at LO and NLO accuracy are compatible with the PHENIX data, 
in contrast to  the conclusion of~\cite{Cooper:2004qe}, in which
feed-down {from $\chi_{c0}$ and $\chi_{c2}$ at $\alpha_S^2$ was incorrectly assumed 
to be the dominant source of $J/\psi$ production}.
This supports the good description of STAR results~\cite{Abelev:2009qa} for the $J/\psi$  differential cross section at mid $p_T$
predicted by the CSM at NLO including leading-$p_T$ $\alpha_s^5$ contributions
(NNLO$^\star$)\cite{Artoisenet:2008fc}. Note that a significantly larger CS yield points to a small impact from
$s$-channel cut contributions~\cite{schannelcut}.

 Even though the NLO is close to the data,
the additional $cg$ contribution (even with a sea-like IC distribution)  improves the agreement.
However, phase-space effects are not properly taken into account in the case 
of $\psi(2S)$ production due to the restriction
$M_{\psi(2S)}=2m_c$. The $\psi(2S)$ case is nevertheless 
encouraging, since it does not involve the uncertainties arising 
from the  extrapolation of the experimental data to the direct yield.
We also give in~\cf{fig:500GeV} our prediction at $\sqrt{s}=500$ GeV for the direct $J/\psi$ and $\Upsilon$
yield for future comparison with  the  data taken this year.

\begin{figure}[!hbt]
\includegraphics[width=0.99\columnwidth]{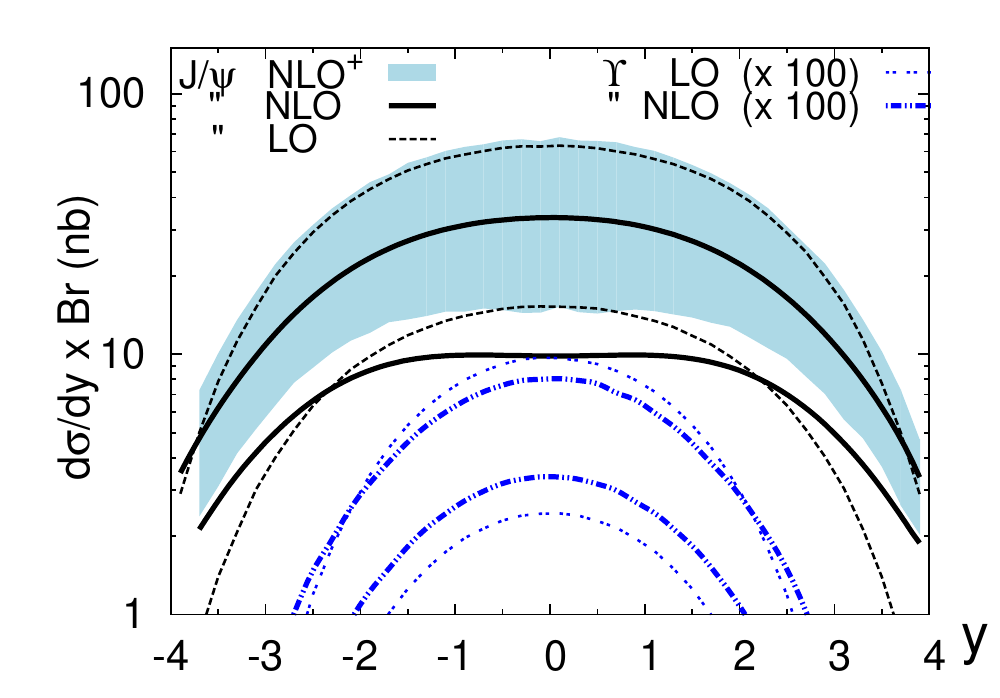}
\caption{$d\sigma/dy \times Br$  for the direct yield of $J/\psi$ and $\Upsilon$ as a function of  $y$
at $\sqrt{s}=500$ GeV for the same parameter ranges as Figs. 1}
\label{fig:500GeV}
\end{figure}

~\\ 

We note that the contribution from $cg$ fusion (the results labeled\footnote{{Notation 
not to be confused with NLO$^\star$ or NNLO$^\star$
which denote real-emission contributions as evaluated in~\cite{Artoisenet:2008fc}}.}  NLO$^+$ were obtained with the sea-like IC from CTEQ 6.5c)
is significant for both $J/\psi$ and $\psi(2S)$ 
production and calls for a deeper analysis.
First, It should be noted that NRQCD 
factorization breaking effects,
such as those arising from the CTM~\cite{Nayak:2007mb,Nayak:2007zb} may impact the low  $p_T$ 
region. Such effects arise from infrared sensitive domains at
NNLO when the 3 heavy quarks have comparable velocities.
A careful study of the CTM is however beyond the scope
of our analysis. Second, to precisely assess
the impact of other choices for the charm distribution, $c(x)$, we have evaluated the fraction of 
$J/\psi$ produced in association with a single $c$-quark relative to the direct yield as a function 
of  $y_\psi$ and for the three models for $c(x)$. These are displayed on \cf{fig:fractionpsic} for
which we have set\footnote{Indeed, for our prediction of the ratio to make sense, the colour singlet contribution 
has to be the dominant one, which can onlybe the case for a rather low charm quark mass such as $m_c  = 1.4$~GeV.} 
 $m_c  = 1.4$~GeV and varied $\mu_F$ and $\mu_R$ within the same values as for 
Figs. 2. 
This clearly confirms the impact of the $cg$ contribution, which ranges from 10 \% up to 45\%
of the direct yield in the case of sea-like $c(x)$.

Note also that at larger $p_T$, we expect significant $\alpha_s^4$ contributions from $cg$ fusion, since they then exhibit a
fragmentation-like topology (\cf{diagrams} (f)). This was studied by Qiao~\cite{Qiao:2003pu} for the Tevatron
using a conventional
$c$-quark distribution, but this evaluation cannot be extended to small $p_T$ where it is infrared divergent.
For the BHPS IC distribution, the $p_T$ distribution
at  large $p_T$ and RHIC energy will show an analogous enhancement as seen at large rapidity in~\cf{fig:fractionpsic}.   This may also impact the $J/\psi$
yield in this region.
In order to assess experimentally the importance of $cg$ fusion,
whether from the usual CSM or from CTM effects, 
the measurement of $J/\psi$ in association
with $D$ meson would be illuminating,  as has been noted in ref. \cite{Artoisenet:2007xi} for $J/\psi + c \bar c$. More accessible is
 the study of the azimuthal correlation of $J/\psi + e$ in the central region by PHENIX and STAR and
of $J/\psi + \mu$ in the forward region by PHENIX. The key signature for such subprocesses is the observation a lepton excess
 opposite in azimuthal angle $\phi$ to the detected $J/\psi$. 
\begin{figure}[!bt]
\includegraphics[width=.99\columnwidth]{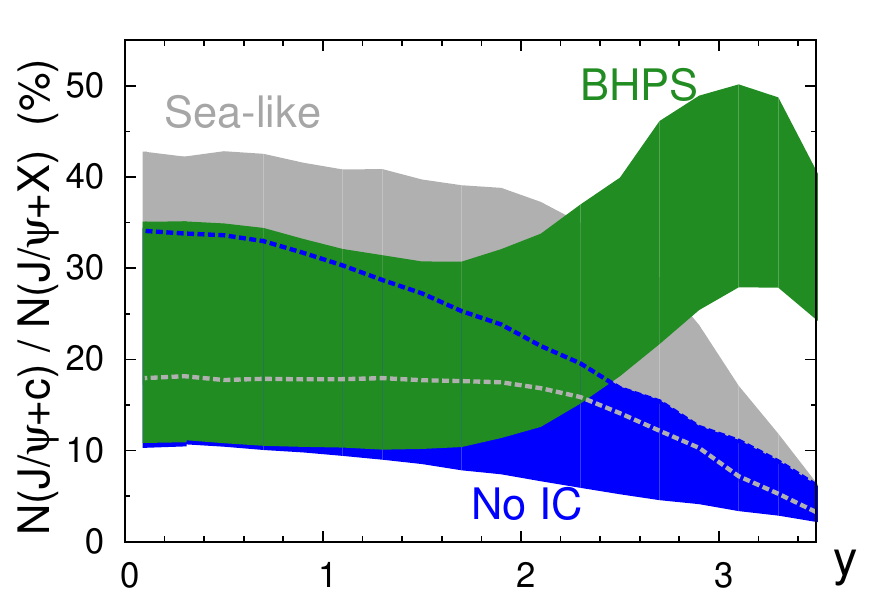}
\caption{Fraction of $J/\psi$ produced in association with a single $c$-quark (via $gc\to J/\psi c$) relative to the direct yield (NLO$^+$) as
a function of  $y_\psi$ and for three models for $c(x)$: without IC (No IC), sea-like and BHPS (see text).}
\label{fig:fractionpsic}
\end{figure}

One can also have, at large rapidity, $(c \bar c ) g \to J/\psi$ contributions to the 
total cross section~\cite{Brodsky:1989ex,Brodsky:1991dj} from the coalescence of the 
charm pair and gluon; in this case the $J/\psi$ acquires the momentum of both the $c$ 
and $\bar c$ quarks from the projectile. Intrinsic charm Fock states such as 
$\vert (c \bar c )_{8_C} (uud)_{8_C}\rangle$ can explain $J/\psi$ and double $J/\psi$ 
production at high $x_F>0.6$ observed in $pA$ and $\pi A$ collisions by the CERN 
NA3 experiment as well and it anomalous $A$ dependence~\cite{Badier:1983dg}.

We now turn to $\Upsilon$ hadroproduction where the  $bg$ fusion processes 
are suppressed by phase-space and by
the $1/m_b^2$ dependence of the $b$-quark content in the proton.
Thus  we have only computed the LO and NLO yield
from $gg$ and $qg$ (see~\cf{fig:xsection} (c)). The predictions are not far from the extrapolation of preliminary data by PHENIX and STAR. In addition, the consistency between  CDF data at the Tevatron at mid and large $p_T$ and the very first NNLO$^\star$ CS analysis~\cite{Artoisenet:2008fc} also suggests that  $\Upsilon$ production can be understood from   perturbative QCD.
We also emphasize here that the rapidity region accessible at RHIC allows for measurements
of $\Upsilon$ production at high $x_F$ very close to 1 where
the intrinsic bottom quark pair can simply coalesce to form a $\Upsilon$ after a single scattering to change its color  in $(b \bar b )_{8_C} + g \to \Upsilon$ in analogy to the large $x_F$ $J/\psi$ production~\cite{Brodsky:1991dj}. It
does not require a third $b$-quark and is thus not suppressed by phase-space effects. 
%One can possibly attribute the  excess
%visible at $|y|\simeq 2$ in the PHENIX measurements to this mechanism.    

We now briefly discuss the production of $J/\psi$
in $pA$ collisions as CS states, likely the dominant mechanism at RHIC energy.
In the central region, the $c \bar c$ pair hadronizes outside the nucleus.
Although the energy loss of a colored object in
cold nuclear matter is limited to be constant, rather than scaling with energy,  by the 
Landau-Pomeranchuk-Migdal effect~\cite{Brodsky:1992nq}, its magnitude per unit of length will be significantly larger for a CO than for a CS
state. The recent observation by STAR~\cite{Abelev:2009qa}
of the non-suppression of $J/\psi$ in Cu-Cu collisions at increasing $p_T$ clearly supports  the hypothesis that
 the $J/\psi$ is produced by a hard subprocess where the $ c \bar c$ is in a colorless state.
The dominant hard QCD subprocess for $J/\psi$ hadroproduction is thus 
a $ 2 \to 2$ reaction in contrast to the feed-down  $ gg \to \chi_{c2} \to J/\psi \gamma$ 
or CO mechanism such as  $gg \to (c\bar c)_{8C} \to J/\psi g$~\cite{Cooper:2004qe}.
 Nuclear shadowing 
should then be implemented along the lines of~\cite{Ferreiro:2008wc}, both for $gg$ and $cg$ part, although the
$c$-quark shadowing is  poorly known.  Thus the  dedicated study
of $J/\psi+c$ in $pA$ collisions could provide a unique way to study such shadowing effects
as well as heavy-quark energy loss.  We also note that
the  yield  from $c g$ subprocesses is expected to have the usual factorizing nuclear dependence
$A^{\alpha(x_2)}$, where $x_2$ is the light-front momentum fraction of the nuclear parton,  in contrast to the factorization breaking behavior $A^{\alpha(x_F)}\sim A^{2/3}$ observed at high $x_F$~\cite{Badier:1983dg,Hoyer:1990us}, explainable
by the coalescence of IC pairs turning into CS pairs after interacting with partons from
the target surface~\cite{Brodsky:1989ex,Brodsky:1991dj,Vogt:1991qd}.

In conclusion, we have carried out the first NLO analysis in the Color-Singlet model of $J/\psi$, $\psi(2S)$
and $\Upsilon$ production at RHIC and have shown that the CS yield is in agreement with the  $p_T$-integrated cross sections
measured by the PHENIX and STAR collaborations.  We have also shown that $c$-quark--gluon fusion is responsible
for a significant, and measurable, part of the yield, and we call for a dedicated measurement to pin down this contribution
and assess the importance of the charm content of the proton. 
Such a study may also shed light on effects due to 
color-transfer  effects beyond the CSM. We  predict a significant excess of the
lepton yield on the ``away" side of the  $J/\psi$ arising from $c$-quark jet and argue that the rapidity dependence 
of this correlation is
strongly sensitive on the specific mechanisms for the creation of charm in the proton. Finally, we have  discussed
the implication of our work on heavy-ion studies.

\vspace*{-.5cm}

\acknowledgments We thank  L. Bland, F. Close,  C. Da Silva, F. Fleuret, R. Granier de Cassagnac, P. Hoyer, 
M. Leitch, J.W. Qiu and T. Ullrich
for useful discussions,
J.~Campbell, F.~Maltoni and F.~Tramontano for their NLO code, as well as
J. Alwall and P. Artoisenet for useful technical advice. This work is supported in part by a Francqui Fellowship
of the Belgian American Educational Foundation and
by the U.S. Department of Energy under contract number DE-AC02-76SF00515.

%%%%%%%%%%%%%%%%%%%%%%%%%%%%%%%%%%%%%%%%%%%%%%%%%%%%%%%%%%%%%%%%%%%%%%%%%%%%

\end{document}